# Finite-volume two-pion energies and scattering in the quenched approximation

CLAUDE W. BERNARD[1] AND MAARTEN F.L. GOLTERMAN[2]

*Department of Physics*
*Washington University*
*St. Louis, MO 63130, USA*

ABSTRACT: We investigate how Lüscher's relation between the finite-volume energy of two pions at rest and pion scattering lengths has to be modified in quenched QCD. We find that this relation changes drastically, and in particular, that "enhanced finite-volume corrections" of order $L^0 = 1$ and $L^{-2}$ occur at one loop ($L$ is the linear size of the box), due to the special properties of the $\eta'$ in the quenched approximation. We define quenched pion scattering lengths, and show that they are linearly divergent in the chiral limit. We estimate the size of these various effects in some numerical examples, and find that they can be substantial.

---

[1] e-mail: cb@lump.wustl.edu
[2] e-mail: maarten@aapje.wustl.edu

## 1. Introduction

Almost all lattice QCD efforts so far have been devoted to the computation of the hadron spectrum and hadronic weak matrix elements. However, much more is known about hadrons than just their spectrum and weak decays, and one would like to derive this rich phenomenology quantitatively from QCD. The example which concerns us here is the computation of pion scattering lengths.

It is difficult to obtain scattering amplitudes directly in lattice QCD, since this would involve an analytic continuation to the physical regime of numerically obtained Euclidean four-point correlation functions. However, Lüscher has shown [1,2] that the energy eigenvalues of states with the quantum numbers of two-particle states in finite volume admit a systematic expansion in $1/L$, where $L$ is the linear dimension of a spatial volume with periodic boundary conditions. In addition, he showed that the coefficients of the powers of $1/L$ in this expansion are related to the infinite-volume elastic scattering phase shifts at values of spatial momenta that occur in the finite volume, for energies below the inelastic threshold. In particular, for the shift $\Delta E$ in the lowest two-particle energy eigenvalue $E$ for spinless bosons with mass $m$ which are at rest, one has

$$\Delta E = E - 2m = -\frac{4\pi a_0}{mL^3}\left(1 + c_1 \frac{a_0}{L} + c_2 \frac{a_0^2}{L^2}\right) + O(L^{-6}), \qquad (1)$$

where $a_0$ is the $l = 0$ scattering length, $c_1 = -2.837297$, and $c_2 = 6.375183$. It is important to note that equations such as these are only expected to be applicable in the domain where the finite-volume corrections to single particle energies, which are exponentially suppressed [3], are indeed small compared to the right hand side of eq. (1).

Relations such as these simplify the numerical effort needed to obtain information on phase shifts enormously. For instance, one can obtain the $l = 0$, $I = 0$ and $I = 2$ pion scattering lengths from eq. (1) by extracting the lowest energy levels in the $I = 0, 2$ channels from the large time behavior of the Euclidean correlation functions

$$
\begin{aligned}
C_{I=0}(t) &= \frac{3}{2}C_{+-}(t) - \frac{1}{2}C_{++}(t), \\
C_{I=2}(t) &= C_{++}(t), \\
C_{+-}(t) &= \langle 0|\pi^+(t)\pi^-(t)\pi^-(0)\pi^+(0)|0\rangle_{\text{con}}, \\
C_{++}(t) &= \frac{1}{2}\langle 0|\pi^+(t)\pi^+(t)\pi^-(0)\pi^-(0)|0\rangle_{\text{con}},
\end{aligned}
\quad (2)
$$

where $\pi^\pm(t) \equiv \int d^3\mathbf{x}\ \pi^\pm(\mathbf{x},t)$ are zero spatial momentum charged pion fields at time $t$. The subscript "con" means that we should exclude graphs where the fields at time $t$ are disconnected from the fields at time 0. The extra factor one half in the last equation comes from the fact that the initial and final $\pi^+\pi^+$ states both need a factor $1/\sqrt{2}$ to be properly normalized because they are states with two identical particles. This method has actually been used in attempts to compute the $I = 2$ scattering length [4,5], and, more recently, the $I = 0$ scattering length (as well as the $I = 2$, the pion-nucleon and the nucleon-nucleon scattering lengths) [6].

However, all these computations were done in the quenched approximation, and it is not clear *a priori* that Lüscher's analysis carries over to the quenched approximation without modification. New infrared divergences occur in quenched QCD in the chiral limit [7,8,9], and modification of the $L$ dependence of equations like eq. (1) might occur, at least for those particles sensitive to the chiral limit. For instance, it was shown that the imaginary part of the quenched $\pi^+\pi^- \to \pi^+\pi^-$ scattering amplitude diverges at threshold (*i.e.*, at zero relative momentum) when calculated using quenched chiral perturbation theory (ChPT) [10]. (For quenched ChPT, see refs. [11,7,9,12]. In this paper, we investigate in more detail the case of pion-pion scattering at threshold in quenched chiral perturbation theory. We will consider only the case of degenerate quark masses.

## 2. Calculation in Quenched ChPT

The correlation functions $C_{+-}(t)$ and $C_{++}(t)$ can be calculated in the ChPT expansion. Formally, with $\chi(t)$ the operator that creates a two-particle state with definite quantum numbers, the Euclidean correlation functions $C = C_{I=0,2}$ can be expressed as

$$
\begin{aligned}
C(t) &= \sum_{|\alpha\rangle} e^{-E_\alpha t} \frac{|\langle 0|\chi(0)|\alpha\rangle|^2}{\langle \alpha|\alpha\rangle} \\
&= \sum_{|\alpha\rangle} e^{-E_\alpha^{(0)} t} \Big[ Z_\alpha^{(0)} + \lambda Z_\alpha^{(1)} + \lambda^2 Z_\alpha^{(2)} \\
&\qquad - (\lambda Z_\alpha^{(0)} E_\alpha^{(1)} + \lambda^2 Z_\alpha^{(0)} E_\alpha^{(2)} + \lambda^2 Z_\alpha^{(1)} E_\alpha^{(1)})t \\
&\qquad + \frac{1}{2}\lambda^2 Z_\alpha^{(0)} (E_\alpha^{(1)})^2 t^2 \Big] + O(\lambda^3),
\end{aligned}
\tag{3}
$$

where $\lambda$ is an expansion parameter, proportional to $m_\pi^2/(16\pi^2 f_\pi^2)$ in unquenched ChPT (we will come to quenched ChPT later). We have used the expansions

$$
\begin{aligned}
E_\alpha &= E_\alpha^{(0)} + \lambda E_\alpha^{(1)} + \lambda^2 E_\alpha^{(2)} + O(\lambda^3), \\
\frac{|\langle 0|\chi(0)|\alpha\rangle|^2}{\langle \alpha|\alpha\rangle} &= Z_\alpha^{(0)} + \lambda Z_\alpha^{(1)} + \lambda^2 Z_\alpha^{(2)} + O(\lambda^3).
\end{aligned}
\tag{4}
$$

The sum over $\alpha$ is a sum over all intermediate states except for the vacuum $|0\rangle$; these states are eigenstates of the complete theory (*i.e.*, to all orders in $\lambda$). From eq. (3) we see that the perturbative corrections to the energy eigenvalues can be extracted from the terms linear in $t$, after taking out the factors $Z_\alpha^{(0)} exp(-E_\alpha^{(0)} t)$. We can calculate the energy eigenvalues in a finite box $L^3$ with periodic boundary conditions by restricting all spatial momenta $\mathbf{p}$ to $\mathbf{p} = 2\pi \mathbf{n}/L$ with $\mathbf{n} \in \mathbb{Z}^3$. Calculating to one loop order (*i.e.*, to order $\lambda^2$) and expanding the result in $1/L$, we should recover the first two terms of eq. (1).

The calculation of the $I = 0$ and $I = 2$ correlation functions at tree level in ChPT is straightforward, giving

$$C_{I=0}(t) = e^{-2m_\pi t} \frac{L^6}{4m_\pi^2} \left[ 1 + \frac{5m_\pi^2}{24 f_\pi^2} \frac{1}{(m_\pi L)^3} + \frac{7m_\pi^2}{4 f_\pi^2} \frac{(m_\pi t)}{(m_\pi L)^3} \right],$$
$$C_{I=2}(t) = e^{-2m_\pi t} \frac{L^6}{4m_\pi^2} \left[ 1 + \frac{m_\pi^2}{12 f_\pi^2} \frac{1}{(m_\pi L)^3} - \frac{m_\pi^2}{2 f_\pi^2} \frac{(m_\pi t)}{(m_\pi L)^3} \right], \quad (5)$$

from which we obtain the well known result

$$a_0^{I=0} = \frac{7}{16\pi} \frac{m_\pi}{f_\pi^2}, \quad a_0^{I=2} = -\frac{1}{8\pi} \frac{m_\pi}{f_\pi^2} \quad (6)$$

($f_\pi = 132\ MeV$ in our conventions), using eqs. (1,3). This tree-level result is also valid in the quenched approximation. If we extended the calculation to one loop, we would find the first two terms of eq. (1), with the infinite-volume one-loop expression for $a_0$ in the $L^{-3}$ term, and the tree-level expression for $a_0$ in the $L^{-4}$ term. For the $L^{-5}$ term one would need to go to higher order in the chiral expansion.

We now wish to determine the quenched approximation version of eq. (1) for $I = 0$ and $I = 2$ pion scattering. We will show that eq. (1) changes drastically. In particular, "enhanced finite-volume corrections" to the infinite-volume result $E = 2m_\pi$ occur at order $L^0 = 1$ and $L^{-2}$. We will first present the calculation, and then discuss this remarkable result.

In order to calculate the quenched one-loop corrections to eq. (5), we employ (Euclidean) quenched ChPT, which was systematically developed in refs. [7,13]. It was shown there that the $\eta'$ meson in the quenched approximation has both single and double pole terms in its propagator:

$$\langle \eta'(p) \eta'(q) \rangle = \delta(p+q) \left( \frac{1}{p^2 + m_\pi^2} - \frac{\mu^2}{(p^2 + m_\pi^2)^2} \right), \quad (7)$$

where $\mu$ is the parameter equivalent to the singlet part of the $\eta'$ mass in unquenched QCD. From experiment, $\mu^2/3 = (500\ MeV)^2$ in full QCD; quenched numerical computations get roughly this value [14,15,16,17]. The vertex $\mu^2$ can actually have momentum dependence [7], which we will ignore for the purpose of this paper. It is the double pole in eq. (7) that will lead to the enhanced finite-volume corrections when an $\eta'$ appears as an internal line in a one-loop diagram contributing to pion scattering.

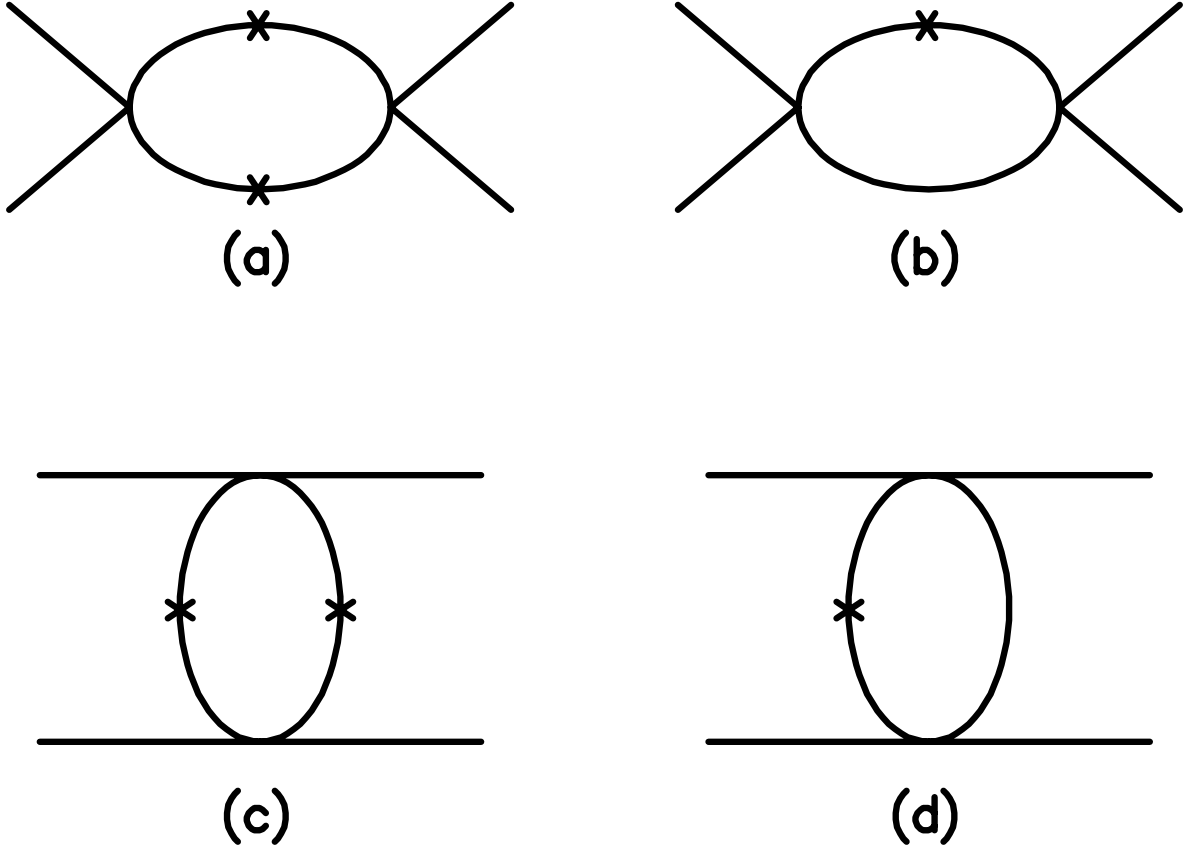

Figure 1. $\mu^2$-dependent s-channel (a and b) and t-channel (c and d) contributions to pion-pion scattering. Incoming (outgoing) particles are on the left (right). An internal line with (without) a cross refers to the double (single) pole term in eq. (7).

There are two types of contributions containing double pole terms. Figs. 1(a) and (b) are the s-channel meson diagrams for the $\pi^+\pi^- \to \pi^+\pi^-$ amplitude ($C_{+-}$) in quenched ChPT, and figs. 1(c) and (d) are the t-channel diagrams for the same amplitude. $C_{++}$ gets t-channel contributions from diagrams like figs. 1(c) and (d) as well as crossed (u-channel) versions. The crosses denote the $\mu^2$-vertex corresponding to the second term in eq. (7), see ref. [7]. Note that any unusual behavior of quenched correlation functions due to the double pole in eq. (7) is always accompanied by $\mu^2$ dependence. This implies that in order to uncover this behavior, we need to calculate the $\mu^2$ dependence of correlation functions. We will do this for $C_{+-}(t)$ and $C_{++}(t)$ to one loop.

These diagrams have been calculated before in Minkowski momentum space (in infinite volume) [10], and it was found that they diverge on shell for $\mathbf{p} \to 0$ as $1/p^3$ (with $\mathbf{p}$ the relative momentum in the center of mass frame); whereas they are perfectly well defined in the Euclidean regime. This seems to indicate that no Hamiltonian formalism can be developed for the quenched approximation and hence that equations like eq. (3) cannot be applied. However we can still parametrize $C(t)$ in Euclidean space as in eq. (3) up to terms linear in $t$, and extract the quantity $\Delta E_\alpha = \lambda E_\alpha^{(1)} + \lambda^2 E_\alpha^{(2)} + \ldots$ from the term linear in $t$ inside the square brackets. (The terms quadratic in $t$, however, will not be of the form indicated in eq. (3) [5]). We will take $E_\alpha^{(0)} + \Delta E_\alpha$ as a definition of the two-particle energy in the quenched approximation. Note that this prescription coincides with the definition used in numerical work [4,5,6].

The calculation of the diagrams in fig. 1 is straightforward. Only the mass term in the chiral Lagrangian contributes to the 4-meson vertices, since in the degenerate mass case the kinetic energy term gives no couplings to the $\eta'$. The $s$-channel contribution to $C_{+-}(t)$ (figs. 1(a) + 1(b)) is

$$C_{+-}^{s,\text{oneloop}} =$$
$$\frac{1}{32 f_\pi^4}\left(\frac{\mu^2}{3}\right)^2 \sum_{\mathbf{k}} \frac{1}{\omega^4(\mathbf{k})} \int dt_1 dt_2 \left(\frac{1}{\omega(\mathbf{k})} + |t_1 - t_2|\right)^2 e^{-2m|t-t_1|-2m|t_2|-2\omega(\mathbf{k})|t_1-t_2|}$$
$$-\frac{1}{16 f_\pi^4}\frac{\mu^2}{3} \sum_{\mathbf{k}} \frac{1}{\omega^3(\mathbf{k})} \int dt_1 dt_2 \left(\frac{1}{\omega(\mathbf{k})} + |t_1 - t_2|\right) e^{-2m|t-t_1|-2m|t_2|-2\omega(\mathbf{k})|t_1-t_2|}, \quad (8)$$

where $\omega(\mathbf{k}) = \sqrt{m_\pi^2 + \mathbf{k}^2}$ and $\mathbf{k} = 2\pi\mathbf{n}/L$ with $\mathbf{n} \in \mathbb{Z}^3$. $t_1$ and $t_2$ are the Euclidean time coordinates of the two vertices in fig. 1, and $\mathbf{k}$ is the spatial loop momentum. The integrals over $t_1$ and $t_2$ can be carried out, and collecting only the terms proportional to $t\, exp(-2m_\pi t)$ we obtain

$$t\, e^{-2m_\pi t}\left[\frac{1}{32f_\pi^4}\left(\frac{\mu^2}{3}\right)^2\left(\sum_{\mathbf{k}\neq 0}\frac{1}{\omega^3(\mathbf{k})}\left(\frac{2\omega^2(\mathbf{k})}{(\mathbf{k}^2)^3}+\frac{1}{2(\mathbf{k}^2)^2}\right)+\frac{45}{32m_\pi^7}\right)\right.$$
$$\left.-\frac{1}{16f_\pi^4}\frac{\mu^2}{3}\left(\sum_{\mathbf{k}\neq 0}\frac{1}{\omega^3(\mathbf{k})}\left(\frac{\omega^2(\mathbf{k})}{(\mathbf{k}^2)^2}+\frac{1}{2\mathbf{k}^2}\right)+\frac{1}{m_\pi^5}\right)\right]. \quad (9)$$

There are also terms proportional to $exp(-2\omega(\mathbf{k})t)$ (multiplied by powers of $t$) arising from eq. (8), corresponding (for $\mathbf{k}\neq 0$) to states with nonzero relative 3-momentum and hence higher energy in eq. (3). For $\mathbf{k}=0$ such terms combine with explicit $exp(-2m_\pi t)$ terms to yield contributions finite at $\mathbf{k}=0$, which have been explicitly separated out in eq. (9). From eq. (9) and from figs. 1(c) and (d) and the corresponding diagrams for $C_{++}(t)$, using eqs. (2,3), correcting for the renormalization factors $Z_\alpha^{(0)}=L^6/(4m_\pi^2)$ (cf. eq. (5)), and noting that all momentum sums are ultraviolet finite, we finally obtain (with $\Delta E_\alpha^{\text{one loop}}=\lambda^2 E_\alpha^{(2)}$, cf. eq. (4))

$$\Delta E_{I=0}^{\text{one loop}}=m_\pi\left[B_0(m_\pi L)\,\delta^2+A_0(m_\pi L)\,\delta\epsilon+O\left(\frac{\epsilon^2}{(m_\pi L)^3}\right)\right],$$
$$\Delta E_{I=2}^{\text{one loop}}=m_\pi\left[B_2(m_\pi L)\,\delta^2+A_2(m_\pi L)\,\delta\epsilon+O\left(\frac{\epsilon^2}{(m_\pi L)^3}\right)\right], \quad (10)$$

where

$$\delta\equiv\frac{\mu^2/3}{8\pi^2 f_\pi^2},\qquad \epsilon\equiv\frac{m_\pi^2}{16\pi^2 f_\pi^2}, \quad (11)$$

and

$$B_0(m_\pi L)=-\frac{20\pi^4 m_\pi}{L^6}\left[\sum_{\mathbf{k}\neq 0}\frac{1}{\omega^3(\mathbf{k})}\left(\frac{2\omega^2(\mathbf{k})}{(\mathbf{k}^2)^3}+\frac{1}{2(\mathbf{k}^2)^2}\right)+\frac{27}{32m_\pi^7}+\sum_{\mathbf{k}}\frac{1}{\omega^7(\mathbf{k})}\right],$$
$$A_0(m_\pi L)=\frac{48\pi^4}{m_\pi L^6}\left[\sum_{\mathbf{k}\neq 0}\frac{1}{\omega^3(\mathbf{k})}\left(\frac{\omega^2(\mathbf{k})}{(\mathbf{k}^2)^2}+\frac{1}{2\mathbf{k}^2}\right)+\frac{1}{m_\pi^5}+\sum_{\mathbf{k}}\frac{1}{\omega^5(\mathbf{k})}\right],$$
$$B_2(m_\pi L)=-\frac{20\pi^4 m_\pi}{L^6}\sum_{\mathbf{k}}\frac{1}{\omega^7(\mathbf{k})},$$
$$A_2(m_\pi L)=\frac{48\pi^4}{m_\pi L^6}\sum_{\mathbf{k}}\frac{1}{\omega^5(\mathbf{k})}, \quad (12)$$

where the $1/\omega^5$ and $1/\omega^7$ terms come from the $t$- (and $u$-) channel diagrams. Note that $\delta$ and $\epsilon$ have to be taken as independent expansion parameters in quenched ChPT [7,12], and that only $\delta$- (*i.e.*, $\mu^2$-) dependent corrections have been included. If the momentum dependence of $\mu^2$ had been included ($\mu^2 \to \mu^2 + \alpha p^2$) then there would be additional terms of order $\alpha\delta\epsilon$ in eq. (10). There are $\delta$-independent contributions as well, but those are proportional to $\epsilon^2$. We will not calculate such terms in this paper. (The $\mathcal{O}(\epsilon^2)$ terms include contributions from the $\Phi_0$-dependent potentials — see ref. [7].) There are no $\delta$-dependent contributions from the one-loop diagrams with a single six-point vertex, because such contributions are absorbed into the renormalization of the pion mass in the tree-level result. This is an example of a phenomenon which occurs quite generally for degenerate quark masses [9].

The complete results to one loop for the energy shifts are then

$$E_{I=0} - 2m_\pi \equiv \Delta E_{I=0} = \Delta E_{I=0}^{\text{tree}} + \Delta E_{I=0}^{\text{one loop}},$$
$$E_{I=2} - 2m_\pi \equiv \Delta E_{I=2} = \Delta E_{I=2}^{\text{tree}} + \Delta E_{I=2}^{\text{one loop}}, \qquad (13)$$

with $\Delta E^{\text{one loop}}$ given by eq. (10) and

$$\Delta E_{I=0}^{\text{tree}} = \frac{-7}{4 f_\pi^2 L^3} ,$$
$$\Delta E_{I=2}^{\text{tree}} = \frac{1}{2 f_\pi^2 L^3} . \qquad (14)$$

We note that $m_\pi$ in eq. (13) and all subsequent equations is the renormalized, finite-volume pion mass, including all one-loop corrections [7].

Some momentum sums that appear in eq. (10) are more singular near $\mathbf{k} = 0$ than in the unquenched case. These momentum sums can expanded in $L^{-1}$ using a result established by Lüscher in ref. [1]:

$$\sum_{\mathbf{k}\neq 0} \frac{f(\mathbf{k}^2)}{(\mathbf{k}^2)^q} \cong L^3 \frac{1}{(2q-2)!} \int \frac{d^3k}{(2\pi)^3} \frac{1}{\mathbf{k}^2} (\Delta_k)^{q-1} f(\mathbf{k}^2) + \sum_{j=0}^{q} \left(\frac{2\pi}{L}\right)^{2(j-q)} f^{(j)}(0) z(q-j), \qquad (15)$$

with integer $q > 0$ and $f^{(j)}(0) = \left(\frac{d}{dk^2}\right)^j f(k^2)|_{k^2=0}$, valid up to corrections vanishing faster than any power of $L^{-1}$ if all derivatives of $f$ are square integrable. $\Delta_k$ is the Laplacian with respect to $\mathbf{k}$ and $z(q) = Z_{00}(q,0)$ is a zeta function, with

$$z(3) = 8.40192397, \quad z(2) = 16.53231596, \quad z(1) = -8.91363292, \quad z(0) = -1. \tag{16}$$

The $t$- and $u$-channel contributions do not contain any singular momentum sums, and instead of eq. (15) we have

$$\sum_{\mathbf{k}} f(\mathbf{k}^2) \cong L^3 \int \frac{d^3k}{(2\pi)^3} f(\mathbf{k}^2) \tag{17}$$

for the nonsingular case, again up to corrections vanishing faster than any power of $L^{-1}$. Using eqs. (15,17), we obtain large $m_\pi L$ approximations for the coefficients $B_{0,2}$ and $A_{0,2}$:

$$\begin{aligned}
B_0(m_\pi L) &\cong -\frac{3z(3)}{8\pi^2} + \frac{3z(2)}{32\pi^2}\left(\frac{2\pi}{m_\pi L}\right)^2 - \frac{5}{12\pi}\left(\frac{2\pi}{m_\pi L}\right)^3 \\
&\quad - \frac{9z(1)}{64\pi^2}\left(\frac{2\pi}{m_\pi L}\right)^4 + \frac{180z(0) - 135}{512\pi^2}\left(\frac{2\pi}{m_\pi L}\right)^6, \\
A_0(m_\pi L) &\cong \frac{3z(2)}{4\pi^2}\left(\frac{2\pi}{m_\pi L}\right)^2 - \frac{1}{2\pi}\left(\frac{2\pi}{m_\pi L}\right)^3 + \frac{3}{4\pi^2}\left(\frac{2\pi}{m_\pi L}\right)^6, \\
B_2(m_\pi L) &\cong -\frac{1}{6\pi}\left(\frac{2\pi}{m_\pi L}\right)^3, \\
A_2(m_\pi L) &\cong \frac{1}{\pi}\left(\frac{2\pi}{m_\pi L}\right)^3.
\end{aligned} \tag{18}$$

The $L^{-3}$ terms in eq. (18) come from the integrals in eqs. (15,17), and the $L^0 = 1$ and $L^{-2}$ terms in the $I = 0$ channel are the enhanced finite-volume corrections that we announced earlier. We also see that the $L^{-4}$ term does not follow the pattern of eq. (1), as can be seen from the fact that it is $\delta$ dependent, whereas the tree-level scattering length is not. There are no enhanced terms in the $I = 2$ channel. Note that the corrections to eq. (18) vanish faster than any power of $L^{-1}$.

The enhanced finite-volume corrections to $\Delta E$ in the quenched approximation make it impossible to define a scattering length *via* eq. (1). The only alternative is to drop the enhanced terms and define the scattering length to be simply $-m_\pi/(4\pi)$ times the coefficient of the $L^{-3}$ term in eq. (1). Using the tree-level result for $a_0^{I=0,2}$ and eq. (11), we then obtain from eq. (10)

$$a_0^{I=0} = \frac{7m_\pi}{16\pi f_\pi^2}\left(1 + \frac{1}{7}\delta + O(\epsilon)\right) + \frac{5\pi}{6m_\pi}\delta^2,$$
$$a_0^{I=2} = -\frac{m_\pi}{8\pi f_\pi^2}(1 + \delta + O(\epsilon)) + \frac{\pi}{3m_\pi}\delta^2. \tag{19}$$

Note that even after removing the enhanced finite-volume corrections, these results still diverge in the chiral limit. We have uncovered yet another example of the bad chiral behavior of the quenched theory [7,8,9,12].

For some values of $m_\pi L$ we have computed the values obtained from the definition of the coefficients $A_{0,2}(m_\pi L)$, $B_{0,2}(m_\pi L)$ (eq. (12)) and also from the approximate expressions given in eq. (18); see tables 1 and 2. From these tables, it is clear that, for the smaller values of $m_\pi L$ currently used in numerical computations, the exact expressions will have to be used.

### 3. Numerical Examples

As a "real world" numerical example, we take $m_\pi = 140$ MeV, $f_\pi = 132$ MeV, $\delta = 0.1$, and $m_\pi L = 6$. We get

$$\Delta E_{I=0}^{\text{tree}} = -1.3 \text{ MeV}; \qquad \Delta E_{I=0}^{\text{one loop}} = -0.3 \text{ MeV},$$
$$\Delta E_{I=2}^{\text{tree}} = 0.36 \text{ MeV}; \qquad \Delta E_{I=2}^{\text{one loop}} = -0.07 \text{ MeV}. \tag{20}$$

Note that quenched ChPT appears to be working reasonably well, with ~25% one-loop corrections to tree-level values. However, because of the enhanced finite-volume corrections, quenched ChPT breaks down for larger $L$. For example, for $m_\pi L = 8$ we find a 63% one-loop correction in the $I = 0$ channel; for $m_\pi L = 12$, a 240% correction. In addition, as $L$ gets smaller the sums in eq. (12) get large contributions from small **k** and are much larger than their (already large) asympotic expansions. This means that, as one might expect, quenched ChPT also breaks down for small volumes. For example, while $m_\pi L = 4$ still gives acceptable one-loop corrections, for $m_\pi L = 2.5$ we find a 94% correction in the $I = 0$ channel.

We note also that $\delta$ as large as $\sim 0.18$ is not excluded; $\delta = 0.18$ is what one would get by putting in the full QCD value for $\mu^2$: $\mu^2/3 = (500\ MeV)^2$. Indeed it is possible that the difference between $\delta = 0.18$ and the $\delta \sim 0.1$ found in simulations [14,15,16,17] is simply due to the larger values of $f_\pi$ which accompany the larger quark masses used on the lattice. With $\delta = 0.18$, (and $m_\pi L = 6$), we have $\Delta E_{I=0}^{\text{one loop}} = -1.3$ MeV and $\Delta E_{I=2}^{\text{one loop}} = -0.3$ MeV, again indicating a breakdown of quenched ChPT.

For the results given in eq. (10) to be useful in numerical simulations, the terms linear in $t$ in eq. (3) must not be obscured by terms with higher powers in $t$. Furthermore, one must be able to separate numerically the contribution of the lowest two-pion state (with relative momentum $\mathbf{k} = 0$) from that of the closest excited states (with relative momentum $|\mathbf{k}| = 2\pi/L$). We therefore demand

$$|\Delta E^{\text{tree}}\, t| \ll 1, \tag{21}$$

and

$$\Omega \equiv 2m_\pi t \left( \sqrt{1 + \left(\frac{2\pi}{m_\pi L}\right)^2} - 1 \right) \gg 1. \tag{22}$$

It is not hard to satisfy these conditions in our "real world" example with $m_\pi L = 6$. Putting $|\Delta E^{\text{tree}}\, t| = 0.1$ we find $\Omega_{I=0} = 10$ and $\Omega_{I=2} = 35$. Note that, because $\Delta E^{\text{tree}} \sim L^{-3}$, $\Omega$ actually decreases for smaller $m_\pi L$ if $|\Delta E^{\text{tree}}\, t|$ is held fixed. For example, at $m_\pi L = 4$, $\Omega_{I=0} = 3$.

We can also calculate the size of the one-loop quenched ChPT effects in a recent numerical computation of scattering lengths by Kuramashi *et al.* [6]. They used a $12^3 \times 20$ lattice at $\beta = 5.7$, and considered pion masses of 0.29 (using staggered fermions) and 0.508 (using Wilson fermions), in lattice units. The values of $f_\pi$ were 0.187 and 0.143 respectively [18] (our $f_\pi$ is $\sqrt{2}$ times that of ref. [18]); we take $\delta = 0.1$.

In their numerical computation Kuramashi *et al.* did not include the "double-annihilation" diagram (in the terminology of ref. [5]), which means that in comparing with eq. (10) we should drop those $O(\delta^2)$ terms that come from $s$-channel amplitudes. This leads to a modification of the expression for $B_0(m_\pi L)$ (*cf.* eqs. (10,12)), which now becomes

$$B_{0,\text{ modified}}(m_\pi L) = -\frac{20\pi^4 m_\pi}{L^6} \sum_\mathbf{k} \frac{1}{\omega^7(\mathbf{k})} = B_2(m_\pi L). \qquad (23)$$

For the various one-loop contributions to $\Delta E$ we obtain using the lattice values for $f_\pi$ (in lattice units):

|  | $a\Delta E_{I=0}^{\text{tree}}$ | $a\Delta E_{I=0}^{\text{one loop}}$ | $a\Delta E_{I=2}^{\text{tree}}$ | $a\Delta E_{I=2}^{\text{one loop}}$ |
|---|---|---|---|---|
| $am_\pi = 0.29$ | -0.029 | 0.0002 | 0.0083 | -0.0017 |
| $am_\pi = 0.508$ | -0.050 | 0.005 | 0.014 | 0.001 |

We see that in this case the one-loop corrections are quite small — anomalously so in some cases because of cancellation between the $\delta^2$ and $\delta\epsilon$ terms. (For $\delta = 0.18$, the one-loop terms are less than 20% of the tree-level terms, except for $\Delta E_{I=2}^{\text{one loop}}$ which is of the same size as $\Delta E_{I=2}^{\text{tree}}$ for $am_\pi = 0.29$.) On the other hand, the conditions eq. (21) and eq. (22) are just barely satisfied, if at all, in the range of $t$ (4 to 9) in which they fit. For example, for $am_\pi = 0.29$, $|\Delta E_{I=0}^{\text{tree}} t| = 0.26$ at $t = 9$, and $\Omega = 2.4$ at $t = 4$. For $am_\pi = 0.508$, $|\Delta E_{I=0}^{\text{tree}} t| = 0.45$ at $t = 9$, and $\Omega = 1.8$ at $t = 4$. Since there are six excited states with $|\mathbf{k}| = 2\pi/L$, considerably larger values of $\Omega$ would be needed to be confident that the excited states are not contaminating the results.

Removing the $\delta^2$ terms in eq. (19) that come from $s$-channel contributions, we would obtain for the $I = 0$ scattering length ($a_0^{I=2}$ remains unchanged):

$$a_{0,\text{ modified}}^{I=0} = \frac{7 m_\pi}{16\pi f_\pi^2}\left(1 + \frac{1}{7}\delta + O(\epsilon)\right) + \frac{\pi}{3m_\pi}\delta^2, \qquad (24)$$

For $am_\pi = 0.29$ the $\delta^2$ term leads to a 3% correction, and for $am_\pi = 0.508$ it is 1%. These numbers are 11% and 2% respectively in the $I = 2$ case.

## 4. Origin of the Enhanced Finite-Volume Corrections

It is not difficult to understand the origin of the enhanced finite-volume corrections to eq. (1) in the quenched theory. First, however, we need to have an intuitive picture in a normal theory of how the $L$ dependence of the right hand side of eq. (1) arises within Euclidean relativistic perturbation theory. (See also ref. [5].)

We begin by examining the factors of $L$ and $t$ in the zeroth order contribution to the correlation functions in eq. (2). At this order, we just have a disconnected product of two separate pion propagators from time 0 to time $t$. Since each of the four external fields is being integrated over space, there would be a factor of $L^{12}$ were it not for the fact that the propagator falls exponentially for $|x| > 1/m_\pi$, which forces the start and end of each line to be close to each other in space, and reduces the $L^{12}$ to $L^6$. Furthermore, each pion line contributes a factor of $e^{-m_\pi t}$; together they produce the factor $e^{-2m_\pi t} L^6$ which appears in eq. (5).

At next order, "tree level," we have one interaction vertex. This vertex is integrated over all space-time, but now all 5 points (4 external plus the vertex) must be spatially close to each other and only a $L^3$ survives. The time, $t'$, of the vertex can however be anything as long as $t > t' > 0$, so this gives a factor of $te^{-2m_\pi t}$ — in other words, this is a contribution to the two-particle energy. Relative to the zeroth order term, this contribution is suppressed by $L^{-3}$ and therefore contributes to the first term on the right hand side of eq. (1).

On the other hand, if, for example, $t' > t$, the total time for propagation is now $t+2(t'-t)$, and the contribution is suppressed by an additional factor $e^{-4m_\pi(t'-t)}$. So only times $t' > t$ with $t' - t \sim 1/m_\pi$ contribute, and this region of integration over $t'$ produces just a factor $\sim (1/m_\pi)e^{-2m_\pi t}$, in other words this is a "Z factor" correction in the sense of eq. (3), not an energy contribution.

Now let us go to the one-loop $s$-channel diagram (in the full theory, or any normal theory, like $\phi^4$). Call the times of the two vertices $t_1$ and $t_2$, with $t_1$ the vertex closest to the outgoing ($t$) lines. In space, all six points (4 external plus 2 vertices) must be close ($\sim 1/m_\pi$) to each other, so we start with a factor of $L^3$. Consider the integration over $t_1$ and $t_2$. If $t > t_1 = t_2 > 0$, the integrand goes like $e^{-2m_\pi t}$. However, as $t_1$ moves away from $t_2$, (but still in the order $t > t_1 > t_2$) the integrand is suppressed by an additional factor $e^{-2(\omega(\mathbf{k})-m_\pi)(t_1-t_2)}$, where $\mathbf{k}$ is the momentum mode in the loop. This forces $t_1 - t_2 \sim \tau_k \equiv 1/(\omega(\mathbf{k}) - m_\pi)$. In infinite volume, the lowest scale in the loop would be $\mathbf{k} \sim m_\pi$. For such $\mathbf{k}$, $\tau_k \sim 1/m_\pi$, $t_1 \approx t_2$, and the integration over $t_1$ and $t_2$ produces single factor of $t$. This is the normal one-loop contribution to the energy of order $L^{-3}$ relative to the disconnected diagram. In other words this is just a correction to $a_0$ in the $L^{-3}$ part of eq. (1).

However, in finite volume one must sum over a discrete set of $\mathbf{k}$'s, and a lower scale for $\mathbf{k}$, $1/L$, is available. Indeed, to estimate the difference between the sum over $\mathbf{k}$ and the integral over $\mathbf{k}$, one can simply look at the contribution of the lowest nontrivial momentum modes, $\mathbf{k} \sim 1/L$. (For the contribution of $\mathbf{k} = 0$, see below).

First note that the factor $L^3 \int \frac{d^3\mathbf{k}}{(2\pi)^3}$, which one would have in "infinite volume" is replaced by $\sum_\mathbf{k}$ in finite volume, so the contribution of any single finite-volume mode is down by $L^{-6}$ from the zeroth order term. The modes $\mathbf{k} \sim 1/L$ have have $\omega(\mathbf{k}) - m_\pi \sim L^{-2}$, so $\tau_k \sim L^2$. Now the integration over $t_1, t_2$ from the region $t > t_1 > t_2 > 0$ gives $\sim \tau_k t e^{-2m_\pi t}$. Since $\tau_k \sim L^2$ this gets boosted back to $L^{-4}$ relative to the zeroth order diagram. In other words, the difference between the sum and integral over $\mathbf{k}$ produces the $a_0^2 L^{-4}$ term in the two-particle energy. Contributions where for example $t_2 > t_1$, or $t_1 > t$, *etc.*, are suppressed with no gain in $L$ factors, so these give higher order corrections in $1/L$.

When $\mathbf{k} = 0$, there is no suppression at all for $t > t_1 > t_2 > 0$: $t_1$ and $t_2$ move freely in this range. But then the integration over $t_1, t_2$ gives a factor $t^2$, so this is not a contribution to the one-loop energy (it is an iteration of the tree-level energy shift). "Out of order" contributions are suppressed by $e^{-4m_\pi \tau'}$, where $\tau'$ is the amount of time out of order. These can give $te^{-2m_\pi t}$, but there is no $L$ enhancement, so these are terms of order $L^{-6}$ in the energy.

Note that the $t$- and $u$-channels do not contribute $L^{-4}$ terms in the energy: the internal lines always contribute $e^{-2\omega(\mathbf{k})(t_1-t_2)}$ (no $\omega(\mathbf{k})-m_\pi$ in the exponent). So $\tau_k \sim 1/m_\pi$ always, and there is no boost in the power of $L$ from the integrations over $t_1, t_2$.

The quenched "anomalous" terms are now easy to understand. An "×" ($\mu^2$ vertex) on an internal line can be at any time with no additional suppression as long as $t_1 > t_\times > t_2$. So the integration over $t_\times$ produces an extra factor of $\tau_k$. Therefore each "×" gives two more powers of $L$, starting from the term $L^{-4}$ in the energy. The single-× diagram's contribution to the energy is thus of order $L^{-2}$, and the double-×'s, order 1. Again, there are no such enhanced contributions in the $t$- and $u$-channels.

We conclude this section with the remark that, in the quenched theory, the enhanced terms in $1/L$ are generated exclusively by the lowest modes in momentum space, *i.e.*, modes with $\mathbf{n} \sim \mathbf{1}$, where $\mathbf{k} = 2\pi\mathbf{n}/L$. The sum over $\mathbf{n}$ for the enhanced terms converges well before $\mathbf{n} \sim m_\pi L$. On the other hand, the $L^{-4}$ terms in the energy in a "normal" theory (*e.g.*, full ChPT), which are analogous to the enhanced quenched terms, are generated by all modes up to $\mathbf{n} \sim m_\pi L$, not just by the $\mathbf{n} \sim \mathbf{1}$ modes. This is because the enhancement by $L^2$ from phase space for modes with $\mathbf{n} \sim m_\pi L$ competes with the $L^2$ enhancement from $\tau_k$ for modes with $\mathbf{n} \sim \mathbf{1}$. Indeed, in a normal theory, the sum over all modes with $\mathbf{n} \lesssim m_\pi L$ contributes at order $L^{-3}$ to the energy, and the $L^{-4}$ terms arise from the *difference* between the sum and the integral over such modes.

## 5. Conclusion

We have shown how to adapt Lüscher's analysis of the relation between the finite-volume energy shifts of two-particle states and scattering lengths to the quenched approximation of QCD. These energy shifts are defined directly from the quenched Euclidean correlation functions, in a way consistent with the way it was done in recent numerical computations of two-pion correlation functions. We have calculated the finite-volume energy shifts to one loop for the case of two pions at rest in quenched chiral perturbation theory; our main result is contained in eq. (10)–eq. (14).

The one-loop corrections are very different from those of the full theory. In particular, "enhanced finite-volume corrections" of order 1 and $L^{-2}$ appear, to be compared with the tree-level contributions which are of order $L^{-3}$. This indicates yet another breakdown of quenched chiral perturbation theory (and presumably quenched QCD) in the infrared, and, as in other quantities, this behavior originates from the special role of the $\eta'$ in the quenched approximation. We have examined the size of these effects in some examples, including one in which we take the pion mass, the pion decay constant and the volume from the most recent numerical computation of the energy shifts [18]. The range of applicability of the whole analysis as a function of the volume, the pion mass, and the $\eta'$ mass is also discussed.

Because of the completely different nature of the quenched one-loop corrections (as compared to the unquenched case), and in particular because of their infrared divergent behavior, the quenched two-pion energy shifts will only be close to the full QCD results if the one-loop corrections are small compared to the tree-level terms. A consequence of this is that one can expect agreement of quenched computations with full QCD at best to 25%, which is the size of one-loop corrections to tree-level chiral perturbation theory in the unquenched case. (See for instance a recent compilation of tree-level and one-loop chiral perturbation theory results for pion scattering lengths by Gasser [19].)

Presumably no satisfactory Hamiltonian formulation exists for quenched QCD. Indications in this direction are the enhanced finite-volume corrections uncovered in this paper, as well as the (related) fact that the pion scattering amplitude, continued to Minkowski space, diverges at threshold [10]. This implies that the energy shifts and scattering lengths as defined in this paper from Euclidean correlation functions do not have a direct physical meaning. One may nevertheless hope that the quenched values for these quantities are close to similarly defined quantities in full QCD (at least for some range of parameters), which do have a physical interpretation. As discussed above, a necessary condition for this is that the values of these quantities are given essentially by tree-level chiral perturbation theory. Finally, we would like to remark that a nonperturbative analysis along the lines of ref. [2] does not seem to be possible, since for that approach a Hamiltonian framework is indispensable. The only analytic handle we have on pion physics in the quenched approximation is Euclidean quenched chiral perturbation theory.

## Acknowledgements

We would like to thank Akira Ukawa for a careful reading of the manuscript, and Steve Sharpe and Pierre van Baal as well as Akira Ukawa for useful discussions. This work is supported in part by the Department of Energy under contract #DOE-2FG02-91ER40628. We are grateful to the National Centre for Theoretical Physics, Canberra, Australia, for hospitality while this work was completed.

**Table 1.** $B_0$ and $A_0$ as a function of $m_\pi L$. The third (fifth) column ($B_{0\,\text{exp}}/B_0$ resp. $A_{0\,\text{exp}}/A_0$) gives the ratio of $B_0$ ($A_0$) calculated from eq. (18) and eq. (12).

| $m_\pi L$ | $B_0$ | $B_{0\,\text{exp}}/B_0$ | $A_0$ | $A_{0\,\text{exp}}/A_0$ |
|---|---|---|---|---|
| 1.0 | -3592.1 | 1.0202 | 9363.3 | 0.50044 |
| 1.5 | -315.47 | 0.96660 | 828.75 | 0.50779 |
| 2.0 | -56.287 | 0.89612 | 151.71 | 0.53075 |
| 2.3 | -24.443 | 0.84633 | 67.895 | 0.55550 |
| 2.5 | -14.902 | 0.81036 | 42.560 | 0.57707 |
| 2.7 | -9.4693 | 0.77256 | 27.993 | 0.60252 |
| 3.0 | -5.1370 | 0.71373 | 16.180 | 0.64664 |
| 3.3 | -3.0011 | 0.65514 | 10.180 | 0.69511 |
| 3.5 | -2.1797 | 0.61837 | 7.7905 | 0.72799 |
| 3.7 | -1.6310 | 0.58529 | 6.1387 | 0.76007 |
| 4.0 | -1.1149 | 0.54625 | 4.5037 | 0.80478 |
| 4.5 | -6.7984 | 0.51968 | 2.9774 | 0.86622 |
| 5.0 | -0.48446 | 0.54330 | 2.1613 | 0.91025 |
| 5.5 | -0.39019 | 0.60041 | 1.6717 | 0.93986 |
| 6.0 | -0.34224 | 0.66856 | 1.3502 | 0.95923 |
| 6.5 | -0.31697 | 0.73276 | 1.1238 | 0.97182 |
| 7.0 | -0.30343 | 0.78697 | 0.95588 | 0.98007 |
| 7.5 | -0.29622 | 0.83043 | 0.82635 | 0.98556 |
| 8.0 | -0.29257 | 0.86450 | 0.72343 | 0.98929 |
| 9.0 | -0.29054 | 0.91166 | 0.57054 | 0.99372 |
| 10.0 | -0.29151 | 0.94055 | 0.46300 | 0.99604 |
| 11.0 | -0.29347 | 0.95875 | 0.38388 | 0.99736 |
| 12.0 | -0.29566 | 0.97058 | 0.32374 | 0.99817 |

**Table 2.** $B_2$ and $A_2$ as a function of $m_\pi L$. The third (fifth) column ($B_{2\,\text{exp}}/B_2$ resp. $A_{2\,\text{exp}}/A_2$) gives the ratio of $B_2$ ($A_2$) calculated from eq. (18) and eq. (12).

| $m_\pi L$ | $B_2$ | $B_{2\,\text{exp}}/B_2$ | $A_2$ | $A_{2\,\text{exp}}/A_2$ |
|---|---|---|---|---|
| 1.0 | -1948.2 | 0.0067546 | 4680.4 | 0.016870 |
| 1.5 | -171.08 | 0.022791 | 413.47 | 0.056581 |
| 2.0 | -30.496 | 0.053939 | 75.145 | 0.13134 |
| 2.3 | -13.219 | 0.081819 | 33.310 | 0.19482 |
| 2.5 | -8.0397 | 0.10476 | 20.682 | 0.24433 |
| 2.7 | -5.0891 | 0.13137 | 13.432 | 0.29864 |
| 3.0 | -2.7327 | 0.17835 | 7.5685 | 0.38638 |
| 3.3 | -1.5677 | 0.23358 | 4.6049 | 0.47712 |
| 3.5 | -1.1178 | 0.27459 | 3.4316 | 0.53665 |
| 3.7 | -0.81582 | 0.31845 | 2.6253 | 0.59376 |
| 4.0 | -0.52955 | 0.38828 | 1.8345 | 0.67251 |
| 4.5 | -0.28357 | 0.50926 | 1.1104 | 0.78032 |
| 5.0 | -0.16838 | 0.62524 | 0.73624 | 0.85795 |
| 5.5 | -0.10892 | 0.72620 | 0.52135 | 0.91027 |
| 6.0 | -0.075482 | 0.80712 | 0.38719 | 0.94408 |
| 6.5 | -0.055213 | 0.86787 | 0.29781 | 0.96541 |
| 7.0 | -0.042100 | 0.91130 | 0.23521 | 0.97868 |
| 7.5 | -0.033138 | 0.94130 | 0.18964 | 0.98688 |
| 8.0 | -0.026731 | 0.96153 | 0.15547 | 0.99193 |
| 9.0 | -0.018349 | 0.98377 | 0.10864 | 0.99694 |
| 10.0 | -0.013249 | 0.99325 | 0.079048 | 0.99884 |
| 11.0 | -0.0099145 | 0.99722 | 0.059348 | 0.99956 |
| 12.0 | -0.0076241 | 0.99886 | 0.045700 | 0.99983 |